# Detection of Ferromagnetic Resonance from 1 nm-thick Co

Shugo Yoshii, Ryo Ohshima, Yuichiro Ando, Teruya Shinjo and Masashi Shiraishi [†]

Department of Electronic Science and Engineering, Kyoto University, Nishikyo-ku, Kyoto 615-8510, Japan

[†]Corresponding author (shiraishi.masashi.4w@kyoto-u.ac.jp)

**Abstract**

To explore the further possibilities of nanometer-thick ferromagnetic films (ultrathin ferromagnetic films), we investigated the ferromagnetic resonance (FMR) of 1 nm-thick Co film. Whilst an FMR signal was not observed for the Co film grown on a $SiO_2$ substrate, the insertion of a 3 nm-thick amorphous Ta buffer layer beneath the Co enabled the detection of a salient FMR signal, which was attributed to the smooth surface of the amorphous Ta. This result implies the excitation of FMR in an ultrathin ferromagnetic film, which can pave the way to controlling magnons in ultrathin ferromagnetic films.

**Introduction**

Nanometer-thick films, so-called ultrathin films, have been collecting broad attention because of their abundant spintronics nature [1-3]. Chiba and co-workers carried out a pioneering study, in which 1 nm-thick ferromagnetic Co film exhibited sizable modulation of its magnetization under an applied gate voltage [1]. Indeed, the Curie temperature of the ultrathin Co was lowered to ca. 320 K and was modulated by gating, which gave a great impact to the spintronics field. A subsequent study using 0.4 nm-thick ultrathin Co revealed greater potential of magnetization control by combining ionic gating [2]. More recently, studies on the spintronic nature of ultrathin films have been expanded to include nonmagnetic metals, and the discovery of a gate-tunable inverse spin Hall effect in ultrathin Pt (2 nm thickness) has opened the field of tunable spin-orbit interaction (SOI) [3]. A common physics among the magnetization control of ultrathin ferromagnetic films and the tunable SOI in ultrathin nonmagnetic films is a substantial shift of the Fermi level under a strong gate electric field [4,5].

Investigation of spin physics using ferromagnetic resonance (FMR) is pivotal in modern spintronics. It has been used in a wide range of studies from both fundamental and applied physics, including those related to the generation of spin current [6] and a spin-torque diode effect [7]. FMR takes place under the simultaneous application of a static external magnetic field and microwave, which enables uniform rotation of magnetization. The uniform magnetization rotation generates a spin wave, a quantized state of which is magnon. Whilst magnons have attracted tremendous attention in spintronics because of their spin current propagation ability [8], they have also attracted great attention in the field of hybrid quantum systems because of the potential of creating strong coupling states with another magnon [9], a photon [10] and a superconducting qubit [11].

Despite the strong potential of magnons in spintronics and other condensed-matter fields, generation of FMR, of which excitation is indispensable to generate magnons in ferromagnets, in ultrathin ferromagnetic metals is not easy. In fact, the thickness of ferromagnetic metal films often used in FMR studies in spintronics is typically greater than 5 nm, because the surface roughness of the substrate beneath the ferromagnetic metal hampers the observation of a clear FMR signal. However, given the success of the substantial magnetization control in ultrathin Co by gating [1,2], efficient and tunable magnon creation in ultrathin ferromagnetic metals under FMR can provide a path to electric-field control of magnons, because the number of magnons is proportional to the square root of the total magnetization and the coupling strength between magnons and photons is collectively enhanced by square root of number of magnons [12]. For the achievement, the first milestone is excitation of FMR with a sufficiently small resonance field and the half-width at half-maximum in ultrathin ferromagnetic metals, resulting in magnon excitation, which has not been sufficiently achieved. In this study, we report the realization of FMR in a 1 nm-thick Co film, which has not been previously demonstrated, by depositing a Ta buffer layer beneath the Co layer. The key to achieving FMR is utilization of smooth surface of amorphous Ta layer.

**Results**

Figures 1(a) and (b) show sample structures and measuring setup, respectively. We prepared two different types of samples: One is $SiO_2$/Co (type-A), and the other is $SiO_2$/Co/Ta (type-B) (see Fig. 1(a)). In type-A samples, a Co thin film of 1,2,3 or 5 nm thickness was deposited onto a $SiO_2$ (300 nm)/Si substrate using radio-frequency magnetron sputtering. In type-B samples, a Ta buffer layer of 3 nm thickness was deposited beneath the 1 nm-thick Co. FMR measurements were performed using a $TE_{011}$ (transverse electric mode) cavity of an electron spin resonance system. Figure 2(a) shows the FMR spectra of the four type-A samples, where the thickness of the Co film was varied. The resonance field and the linewidth of the spectra changed dramatically with the Co thickness. More importantly, an FMR signal was barely observed when the thickness of the Co was less than 2 nm.

To circumvent the problem of the missing FMR in thin Co films, we introduced Ta as a buffer layer (type-B sample). Amorphous Ta is widely recognized to possess a quite smooth surface [1,2,13], although direct observation of the smooth surface of Ta was difficult without exposing the surface to air in our experimental setup and structural analyses such as x-ray diffraction cannot be implemented due to the thin Ta layer (3 nm). Hence, we expected that the insertion of Ta beneath ultrathin Co enables formation of a flat and continuous Co film with nm-thick, yielding a sharp FMR spectrum from the Co. In fact, Chiba and co-workers observed the anomalous Hall effect from 0.4 nm-thick Co grown on 3 nm-thick Ta, which is compelling evidence for the formation of a continuous film [1,2]. Figure 2(b) shows a comparison of the FMR spectra of the Co (1 nm) and Co (1 nm)/Ta (3 nm) samples. The difference in the FMR spectra is clearly discernable, and the FMR spectrum of the Co (1 nm)/Ta (3 nm) sample is, in fact, quite obvious. Given that amorphous Ta possesses a smooth surface, the clear FMR spectrum of the Co is attributed to the insertion of the Ta buffer layer with a smooth surface beneath the 1 nm-thick Co. Figure 2(c) shows the resonance fields, the half-width at half-maximum ($\Delta\mu_0 H$) of the FMR spectra, and the Gilbert damping constant $\alpha$ as functions of the Co layer thickness of the type-A and -B samples. The resonance field and the $\Delta\mu_0 H$ were obtained by deconvolution of the integral form of the FMR spectra (for more detail, see Methods and Supplementary Information). The Gilbert damping constant $\alpha$ was calculated using the following equation,

$$\alpha = \frac{\Delta\mu_0 H \cdot \gamma}{2\pi f_{\text{res}}}, \qquad (1)$$

where $\gamma$ is the gyromagnetic ratio of Co and $f_{\text{res}}$ is the applied microwave frequency [14]. Whilst the resonance field, the half-width at half-maximum of the FMR spectra, and the Gilbert damping constant monotonically increased with decreasing the Co thickness, the 1 nm-thick Co layer overlying a Ta buffer layer exhibited noticeable suppression of them. Thus, it is corroborated that the insertion of an amorphous Ta layer beneath an ultrathin Co layer is an efficient approach to excite FMR in the 1 nm-thick Co.

Since the thickness of the amorphous Ta layer in the previous studies was fixed at 3 nm [1,2], the thickness of the Ta buffer layer was varied from 2 to 5 nm in increments of 1 nm and the FMR of the 1 nm-thick Co on Ta buffer layers of various thickness was measured. The FMR spectra of the Co layers with Ta buffer layers of various thickness and the Gilbert damping constant of each sample are shown in Figs. 3(a) and 3(b), respectively. Neither the FMR spectra nor the magnitude of the Gilbert damping constant is dependent on the Ta thickness, which suggests that the insertion of a 3 nm-thick Ta buffer layer is sufficient to induce formation of a flat Co layer. Notably, an FMR signal was not observed from the Co (1 nm)/ Ta (1 nm) sample (see Supplementary Information), which directly indicates that an excessively thin Ta buffer layer does not allow exciting the FMR in a 1 nm-thick Co layer.

To better understanding the aforementioned phenomena, we prepared a Co (1 nm)/ Ⓞ /Ta (3 nm) sample, where the surface of the Ta was intentionally oxidized (Ⓞ denotes that the sample was exposed to air at this sample fabrication step). Figure 4(a) shows a comparison of the FMR spectra of the Co (1 nm)/Ta (3 nm) and Co (1 nm)/ Ⓞ /Ta (3 nm) samples. A substantial difference in the FMR spectra is observed; an FMR signal was not observed from the Co (1 nm)/ Ⓞ /Ta (3 nm) sample albeit a 3 nm-thick Ta buffer layer was introduced. The lack of an FMR signal from the Co (1 nm)/ ⓪ /Ta (3 nm) sample is thus attributed to the oxidized surface of the Ta. As aforementioned, the insertion of an amorphous Ta layer beneath the 1 nm Co facilitates the formation of a flat and continuous Co film because the surface of the amorphous Ta is smooth. Meanwhile, the surface roughness of the oxidized 3 nm-thick Ta was measured to be almost the same as that of the $SiO_2$ substrate; a roughness of ca. 1 nm was observed for both samples by atomic force microscopy (AFM), whilst the grain sizes of these two samples slightly differed (see Figs. 4(b)). Here, to note is that surface of amorphous Ta cannot be measured by AFM without exposing the sample surface to air in our measuring setup. Hence, we deduce that the surface of the oxidized Ta loses sufficient smoothness, which can hinder the formation of a flat and continuous 1 nm-thick Co film. These results unequivocally rationalize that inserting a Ta buffer layer and maintaining its smooth surface play crucial roles in the growth of an ultrathin Co layer that can generate salient FMR.

**Discussion**

The upshift of the resonance field as a function of the Co thickness in type-A samples (see Figs. 2 (a) and (c)) is attributed to a decrease of the total magnetization. Chiba and co-workers observed strong suppression of magnetization in 1 nm-thick Co, resulting in a substantial decrease of the Curie temperature [1]. Hence, the upshift of the resonance field is due to weaker magnetization of the ultrathin Co, consistent with the previous study [1]. Meanwhile, the missing FMR spectra is attributed to roughness of the Co film. We used $SiO_2$/Si substrates, of which surface is not sufficiently smooth. Indeed, it was previously found that the FMR spectra of $Ni_{80}Fe_{20}$ (Py) were strongly dependent on the substrates and that the FMR linewidth of Py grown on a $SiO_2$ substrate was greater than the linewidths of Py grown on yttrium-iron-garnet and non-doped diamond substrates [15]. The broader FMR spectrum of Py on a $SiO_2$ substrate is ascribed to the roughness of the a $SiO_2$ substrate, which hampers isotropic magnetization and uniform magnetization precession under FMR. The results obtained in the present study are consistent with those reported in the literature [15].

In chronicle of FMR studies of thin ferromagnetic films, a couple of studies were implemented about detection of FMR spectra from ultrathin Co [16,17] almost two decades ago. The thickness of the Co films in those previous studies is comparable to that in our present study. Meanwhile, the resonance field and the FMR linewidth in those studies were roughly 500 mT and 60 mT (note that the linewidth of 60 mT is equivalent to the half-width at half-maximum of roughly 70 mT) [16], respectively (the similarly large resonance field was also reported in the literature [17], whereas the linewidth was not discussed in the study). Such the large resonance field and the broad FMR linewidth are attributed to a fact that the ferromagnetism was quite weak and the magnetization precession under FMR was not uniform, i.e., the quality of the Co was poor. Furthermore, the FMR spectrum exhibited the single branch only when the thickness of the Co was greater than 2 nm in that study [16]. Given that the resonance field and the half-width at half-maximum of the FMR spectra in our study are smaller and roughly 80 mT and 15 mT, respectively, and that the single FMR spectrum can be seen from the 1 nm-thick Co, the ultrathin Co used in our study simultaneously possesses sufficiently strong ferromagnetism and uniform FMR unlike in the previous studies, i.e. we experimentally demonstrated that the quality of the ultrathin Co with a Ta buffer layer is sufficiently good and the ultrathin Co is quite available for future magnon spintronics and quantum hybrid systems.

In summary, we achieved FMR excitation from an ultrathin Co film with a thickness of 1 nm by inserting an amorphous Ta buffer layer. The smoothness of the amorphous Ta played a crucial role in a formation of a flat and ultrathin Co layer. These findings provide a new pathway to the electric-field control of magnons using ultrathin ferromagnetic metals.

**Methods**

In type-A samples, a Co thin film of 1,2,3 or 5 nm thickness was deposited using radio-frequency magnetron sputtering. The base pressure of the sputtering system was kept to be lower than $2.5 \times 10^{-5}$ Pa, and the flow rate and partial pressure of the Ar gas were set to be 5 sccm and 0.5 Pa, respectively; and the sputtering temperature was room temperature (RT). The deposition rate of the Co was 0.8 nm/min. A $SiO_2$ layer (10 nm) was deposited onto the Co film to prevent from oxidation of the Co layer. In type-B samples, a Ta buffer layer of 3 nm thickness was deposited, where the deposition rate of the Ta was 3.83 nm/min. The substrate was the same as that used for the type-A samples. During the deposition of Co and Ta, the sample holder was rotated at 20 rpm.

FMR measurements were performed using a TE$_{011}$ (transverse electric mode) cavity of an electron spin resonance system (JEOL JES-FA 200); an external magnetic field under microwave irradiation was applied parallel to the sample plain, and the microwave frequency and power were set to be 9.12 GHz and 10 mW, respectively (see Fig. 1(b)). All of the FMR measurements were carried out at RT.

Deconvolution of the FMR spectra was carried out by using the integral form of the FMR spectra. In addition to the Lorentzian (symmetric) component, an asymmetric component was taken into account. The fitting function used is $F(\mu_0 H) = A_{\text{sym}} \frac{(\Delta \mu_0 H)^2}{(\mu_0 H - \mu_0 H_{\text{res}})^2 + (\Delta \mu_0 H)^2} - 2A_{\text{asym}} \frac{\gamma(\mu_0 H - \mu_0 H_{\text{res}})}{(\mu_0 H - \mu_0 H_{\text{res}})^2 + (\Delta \mu_0 H)^2} + a(\mu_0 H) + b$, where $A_{\text{sym}}$ and $A_{\text{asym}}$ are symmetric and asymmetric components, respectively, and, $a$ and $b$ are constants.

**Acknowledgement**

This work is supported in part by a Grant-in-Aid for Scientific Research (S), "Semiconductor spincurrentronics" (No. 16H06330) and Izumi Science and Technology Foundation. The authors thank Prof. Daichi Chiba of Osaka University, Japan, for his fruitful discussion and suggestions about the growth of an amorphous Ta layer.

**Author contributions**

M. S., Y. A. and R. O. conceived the experiments. S.Y. fabricated samples, collected data and analyzed results. R.O. helped in the experiments. S.Y., R.O. and M.S. wrote the manuscript. All authors discussed the results.

**Competing interest**

The authors declare no competing interests.

**Additional information**

**Supplementary information** is available for this paper at XXX.

**Correspondence** and requests for materials should be addressed to M.S.

**Figure captions**

**Figure 1.** (a) Schematics of the sample structures of the type-A and the type-B samples. The thickness of the Co ($t_{Co}$) in the type-A samples was 1, 2, 3, or 5 nm. The samples were capped with 10 nm-thick $SiO_2$ to prevent oxidization of the Co. (b) Schematic of the setup used for FMR measurements.

**Figure 2.** (a) FMR spectra of sample-A Co of 1, 2, 3, and 5 nm in thickness. (b) Comparison of the MR spectra of the Co (1 nm) and Co (1 nm)/Ta (3 nm) samples. (c) The Co thickness dependence of the resonance field $\mu_0 H_{res}$ (upper panel), the half-width at half-maximum of the FMR spectra $\Delta H$ (middle panel), and the Gilbert damping constant $\alpha$ (lower panel).

**Figure 3.** (a) FMR spectra from the Co (1 nm)/Ta ($t_{Ta}$ nm) samples. The thickness of the Ta ($t_{Ta}$) was changed from 2 to 5 nm. (b) The Ta thickness dependence of the Gilbert damping constant $\alpha$.

**Figure 4.** (a) Comparison of the FMR spectra of the Co (1 nm)/Ta (3 nm) and Co (1 nm)/ⓞ/Ta (3 nm) samples. (b) Atomic force microscopic views of the surfaces of the $SiO_2$ substrate and the oxidized Ta.

**Figures**

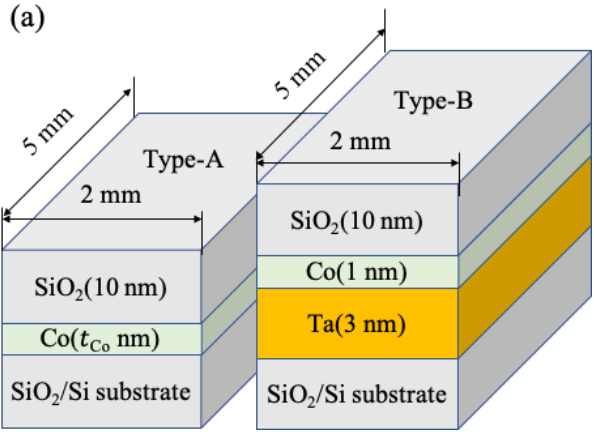

Fig. 1(a) Yoshii et al

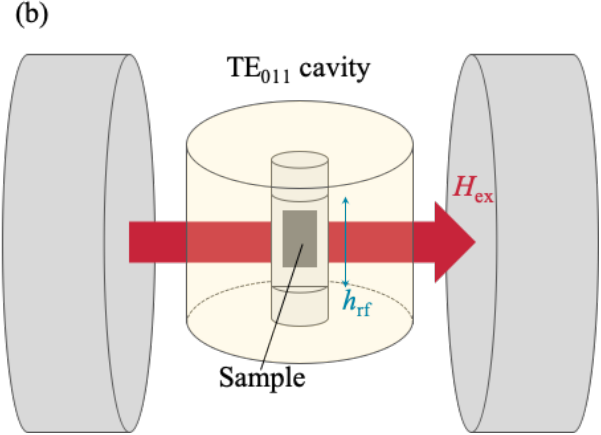

Fig. 1(b) Yoshii et al

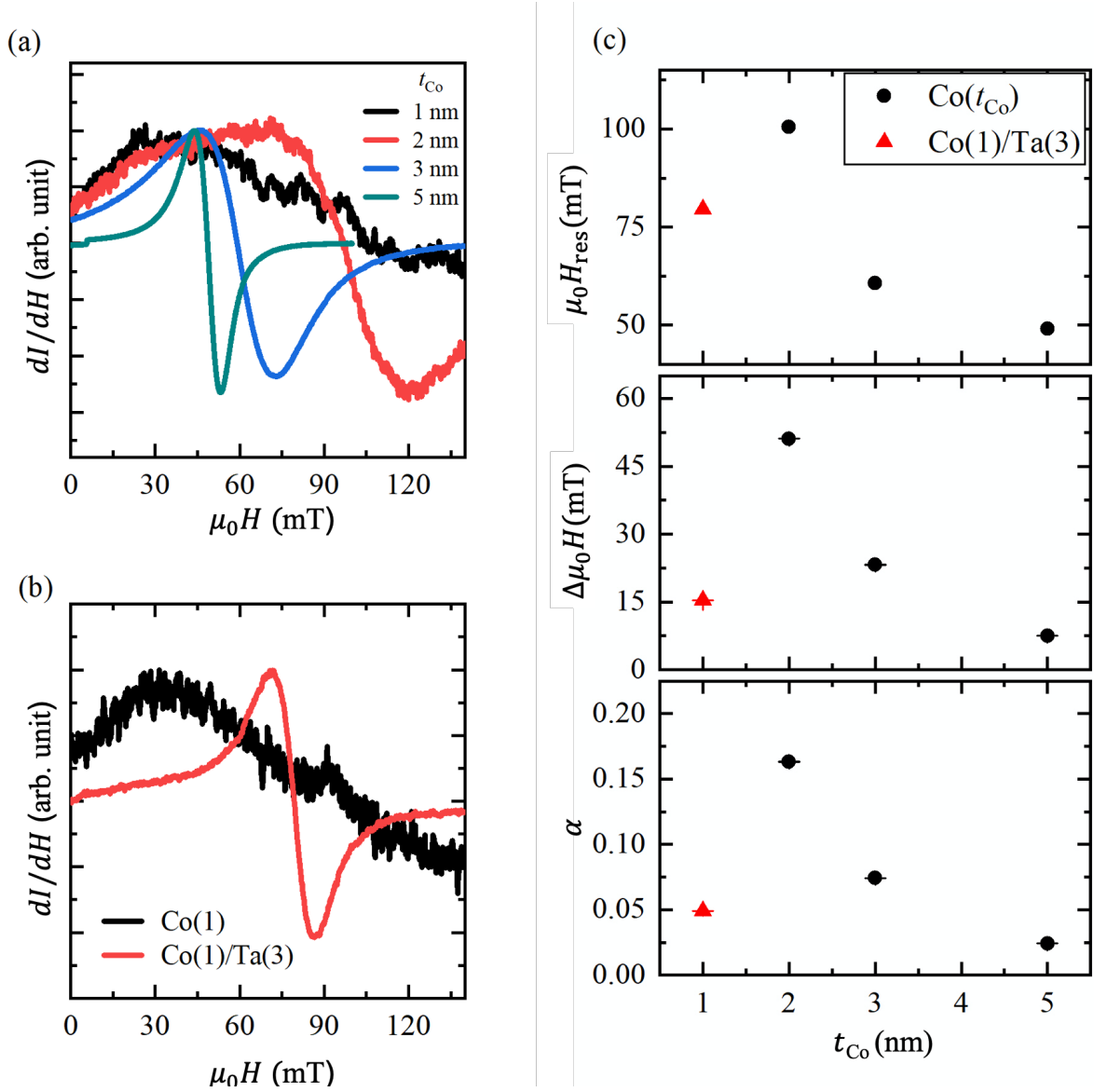

Fig. 2(a)(b)(c) Yoshii et al.

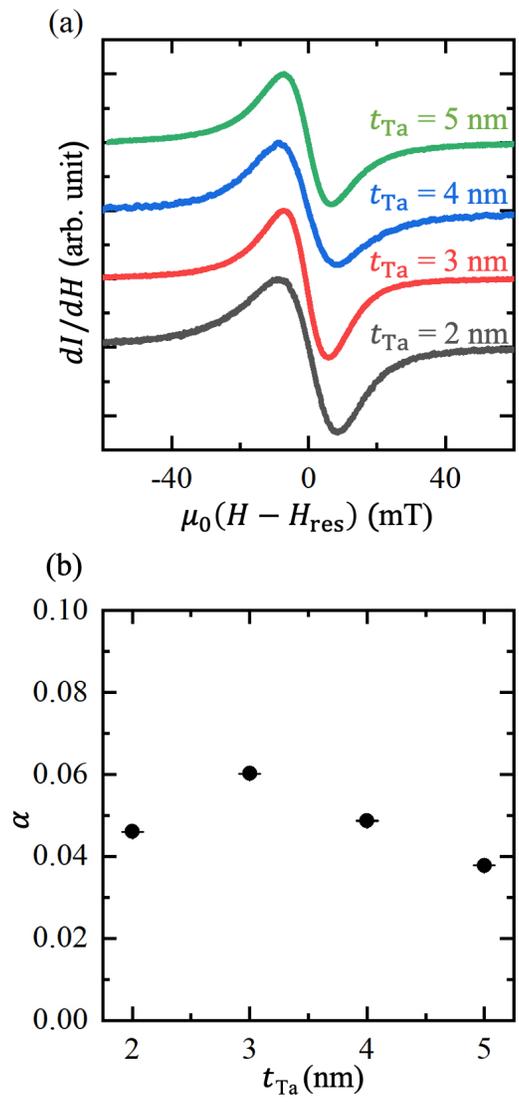

Fig. 3(a)(b) Yoshii et al.

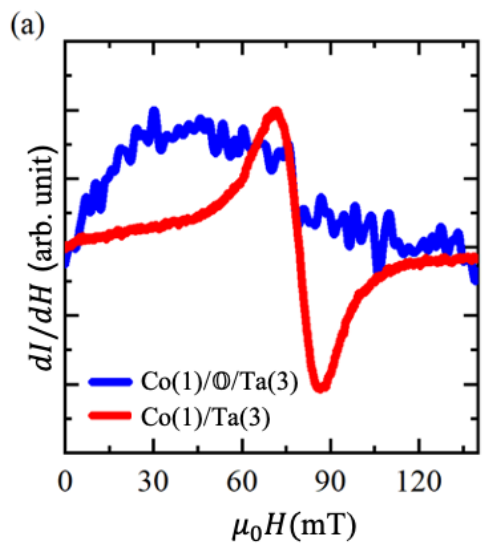

Fig 4(a). Yoshii et al.

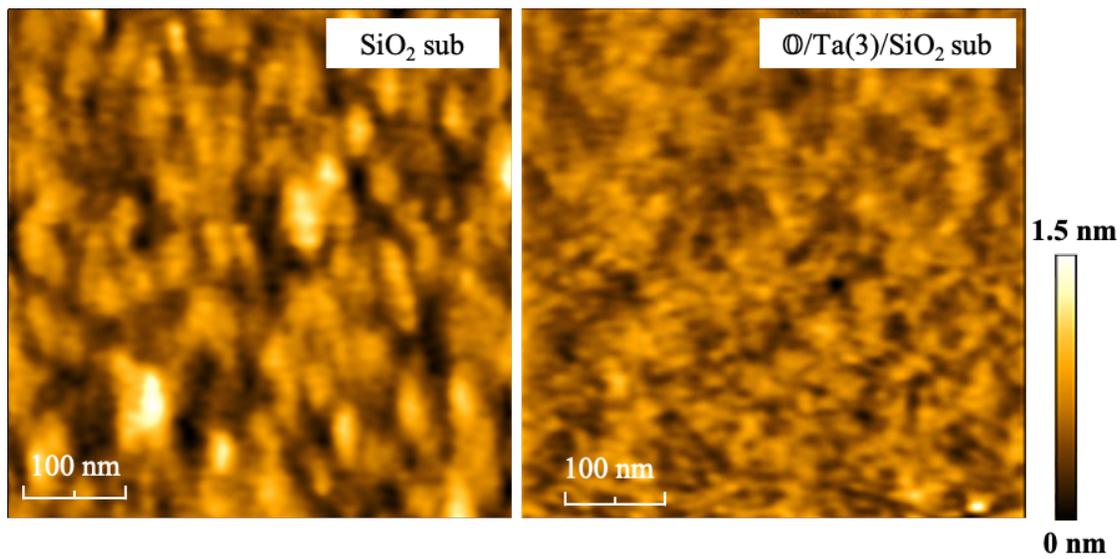

Fig 4(b) Yoshii et al